\documentclass{article}

\usepackage{arxiv}

\usepackage[utf8]{inputenc} % allow utf-8 input
\usepackage[T1]{fontenc}    % use 8-bit T1 fonts
\usepackage{hyperref}       % hyperlinks
\usepackage{url}            % simple URL typesetting
\usepackage{booktabs}       % professional-quality tables
\usepackage{amsfonts}       % blackboard math symbols
\usepackage{amsmath}
\usepackage{amssymb}
\usepackage{nicefrac}       % compact symbols for 1/2, etc.
\usepackage{microtype}      % microtypography
\usepackage{lipsum}
\usepackage{graphicx}
\graphicspath{ {./images/} }
\usepackage[authoryear,round]{natbib}

\title{Experimental investigation of a multi-buoy
cooperative point-absorber wave energy converter}

\author{
  Herman Martens Meyer \\
  Department of Mathematics \\
  University of Oslo \\
  Oslo, Norway \\
  \texttt{hermam@math.uio.no} \\
  \And
  Leif Arne Tønnessen \\
  Concrest Energy \\
  Bærum, Norway \\
  \And
  Olav Gundersen \\
  Department of Mathematics \\
  University of Oslo \\
  Oslo, Norway \\
  \And
  Atle Jensen \\
  Department of Mathematics \\
  University of Oslo \\
  Oslo, Norway \\
}

\begin{document}
\maketitle

%% Abstract
\begin{abstract}

This study presents a proof-of-concept experimental investigation of a multi-buoy cooperative point-absorber wave energy converter (WEC). The proposed concept consists of an array of surface-penetrating buoys connected through a shared closed-loop hydraulic power take-off (PTO) system. Energy is extracted through the collective motion of the buoy array, where pressurised flow generated by individual buoys drives a turbine within the hydraulic circuit. A 1:40 scale model was tested in the wave tank facilities at the University of Oslo. Experiments with regular and irregular long-crested waves at two different incident angles were conducted to assess power absorption, wave period response, and interaction effects. Two array configurations were investigated: an eight-buoy array with an axis-to-axis spacing of 1.5 buoy diameters, and a four-buoy array with a 3.0 diameter spacing. Although piston head leakage affected the power measurements, our results demonstrate that the WEC absorbs incoming wave energy and produces measurable power. The eight-buoy configuration achieved the highest power output per buoy compared to the four-buoy configuration, but exhibited increased sensitivity to wave period and wave heading, due to buoy-buoy interactions, such as collisions. This study highlights buoy count and internal buoy spacing as key design parameters for cooperative point-absorber wave energy systems. The results indicate that higher buoy counts enhance hydraulic cooperation, and increased buoy spacing improves robustness to wave heading and reduces destructive interaction effects. We also suggest that a lower system inertia can improve responsiveness to shorter waves. These insights provide a foundation for further optimisation and future full-scale development.

\end{abstract}

%% Use \section commands to start a section
\section{Introduction} \label{sec:introduction}

% \textcolor{orange}{\textbf{Research question: } How does this new concept for wave energy conversion respond to different wave conditions, ranging from regular waves to more complex and realistic wave conditions? How is wave direction and buoy separation impacting the power production and how can this inform the WECs of tomorrow?}

As global energy demand is rapidly increasing \citep{eia_international_2019} and global consensus is leaning towards greener alternatives, wave energy conversion stands out as an untapped resource with significant theoretical potential. With oceans covering more than 70\% of the Earth and surface waves estimated to contain power on the order of 1-10 TW \citep{panicker_power_1976,10.1115/OMAE2010-20473,gunn_quantifying_2012}, wave energy remains a substantially underdeveloped field. While solar and wind energy have converged on optimal designs, wave energy technologies remain unsettled, with many device and power-take-off (PTO) systems proposed \citep{guo_review_2022}.

Unlike solar and wind, waves offer high availability (90\% \citep{pelc_renewable_2002,guo_review_2022}). A promising use case is co-locating wave energy conversion systems with wind or solar parks to smooth power output and share cost for infield infrastructure and services \citep{sewter_co-location_2025,wang_power_2025}. Waves naturally lag behind the wind, and when combined, can reduce the variability of wind power \citep{astariz_output_2016}.

Since the early concepts of \cite{salter_wave_1974} and \cite{budal_resonant_1975}, wave energy converter (WEC) development has led to a wide range of device types. These are commonly grouped into oscillating water column devices, overtopping converters, and oscillating body systems \citep[see reviews by][]{falcao_wave_2010,arrosyid_recent_2025}. The latter category spans a wide range of floating or submerged structures that move with the wave motion. Within this category are point absorbers, which extract energy from motions at a single point and are small compared to the incoming wave \citep{falnes_review_2007}. \cite{guo_review_2022} provides a comprehensive review of point absorber WECs. Attenuators form another type of oscillating body system, mainly consisting of multiple floating structures aligned parallel to the incoming waves, where energy is extracted from the relative motion between the bodies \citep{yemm_pelamis_2012,stansby_m4_2024}. Both concepts rely on oscillatory body motion, and the concept presented later shares several hydrodynamic mechanisms with both. While point absorbers operate mainly in the heave mode \citep{guo_review_2022}, it has been shown that a combination of multiple degrees of freedom (DOFs) could increase power production \citep{sergiienko_feasibility_2018}. Attenuators, such as the M4 \citep{stansby_capture_2015}, absorb power from the pitch motion between the different elements of the WEC and utilise that different parts of the elongated body that are located in different wave phases. Although power absorption can increase by combining multiple DOFs, the coupling between these modes can significantly impair power absorption if not properly tuned \citep{guo_hydrodynamic_2018,tran_design_2022}.

This interaction between different DOFs can also influence the diffracted and radiated wave field from each WEC. Many WEC concepts are envisioned to be deployed in large wave energy parks with multiple elements, and their interactions introduce a whole new level of complexity. This park effect is studied both analytically, numerically \citep[see parts of][]{goteman_advances_2020} and experimentally \citep{vervaet_experimental_2022}. More relevant for the present study, the spacing between point absorbers is investigated from $1.25D-20D$ in the literature \citep[see review by][]{vervaet_experimental_2022}, and it has been shown to lead to significant alterations in the wave field for more than $40D$ \citep{babarit_impact_2010}, with $D$ being the point absorber diameter. This park effect can lead to both constructive and destructive interaction effects within the array \citep{babarit_park_2013}.

Experimental model testing plays a crucial role in understanding hydrodynamic effects and in validating numerical models prior to full-scale development \citep{vervaet_experimental_2022}. Numerous studies have used downscaled models to investigate device performance, mooring behaviour, survivability, and energy capture across various sea states \citep{brito_experimental_2020,stansby_m4_2024,liao_modelling_2023,xu_experimental_2019,kofoed_prototype_2006,ning_hydrodynamic_2016,shahroozi_experimental_2022}. Despite this progress, there remains a need for experimental data on new WEC concepts. There is also a need to identify and develop wave energy solutions that are both affordable and reliable.

%the development of WECs, allowing for improvement in physical understanding of performance as a bridge from model to full-scale testing and to confirm or calibrate the numerical models \citep{vervaet_experimental_2022}. Several studies on WECs are related to this numerical/experimental connection \citep{sun_linear_2017,stansby_hydrodynamics_2020,paduano_experimental_2020} and there is a lot related to downscaled model testing \citep{brito_experimental_2020,stansby_m4_2024,liao_modelling_2023,xu_experimental_2019,kofoed_prototype_2006,ning_hydrodynamic_2016,shahroozi_experimental_2022} both with the goal of investigate potential power output, but also to model mooring, survivability and other aspects related to full-scale development.

In this article, we present a novel wave energy conversion concept that combines the point-absorber and attenuator principles. This device is tested in the new wave tank at the University of Oslo under a range of regular and irregular wave conditions to estimate power output across different seas and to assess device performance across wave conditions. The purpose of this study is to conduct a proof-of-concept study of the WEC, and we will focus on birds-eye-view characteristics such as power capture, wave period response, and interaction effects. This, along with the details of the 1:40 model of the WEC, is presented in Sec.~\ref{sec:concept}. In Sec.~\ref{sec:validation_and_setup} we present the wave tank facilities and the experimental setup. In Sec.~\ref{sec:results}, we present and discuss the experimental results as well as some suggestions for future design improvements and further testing before conclusions are drawn in Sec.~\ref{sec:conclusion}.

% \textbf{Et viktig resultat for oss er at vi kan få en jevn CWR, noe jeg ikke har sett enda. M4 finner også optimal CWR/power når $\lambda \approx L_{installalasjon}$.}

\section{Theory and performance measures} \label{sec:theory}

%\textbf{Kanskje dette skal gå i et senere kapittel?}

%In quantifying the efficiency and performance of the WEC array, we conduct measurements of flow and pressure generated by the system for a range of different regular long-crested waves as well as irregular waves.

As will be described in Sec.~\ref{sec:concept}, we utilise the measured flow rate, $Q$, and the pressure difference between the high and low pressure side of the system, $\Delta p$, to obtain the absorbed power by the WEC, $P_{\text{abs}} = Q \Delta p$. For the subsequent analysis, we assume linear waves as we mainly operate with low wave steepness, $ak<0.1$, where $k = 2\pi/\lambda$ is the wave number and $a$ the wave amplitude. In terms of providing device performance, we utilise some simple relations.

Incident wave power can be expressed as power per unit of crest length (W/m), and for linear waves, we have

\begin{equation}
    J = \rho g \int_0^\infty S(f)c_g(f)df,
\end{equation}
where $\rho$ is water density, $g$ gravitational acceleration, $S(f)$ is the spectral density of the incident wave, and $c_g$ is the group velocity for intermediate-depth waves, which is the case in the present study \citep{arean_integrated_2017,choupin_integration_2022}. For monochromatic waves, this simplifies to

%, or as

% \begin{equation}
%     J = \frac{\rho g^2}{64 \pi}H_s^2T_e\text{tanh}(kh)\left(1+\frac{2kh}{\text{sinh}(2kh}\right)
% \end{equation}
% %

\begin{equation}
    J = \frac{1}{2}\rho g a^2 c_g,
    \label{eq:J_reg}
\end{equation}
where $c_g$ is the group velocity, defined as

\begin{equation}
    c_g = \frac{1}{2} \left(1+ \frac{2kh}{\text{sinh}(2kh)}\right)\frac{\omega}{k}   
    \label{eq:group_vel}
\end{equation}
and $\omega ^2 = gk~\text{tanh}(kh)$ is the dispersion relation. To quantify the performance of WECs, the capture width, $L_c$ \citep{budal_resonant_1975}, is a metric which compares the absorbed power to the available incident wave power per unit crest length, such that

\begin{equation}
    L_c = P_{\text{abs}} / J.
    \label{eq:capture_width}
\end{equation}
Further, a non-dimensional quantity is obtained by normalising the capture width by a characteristic length scale, $B$, often the device width, yielding the capture width ratio, defined as

\begin{equation}
    \text{CWR} = L_c / B,
    \label{eq:CWR}
\end{equation}
by \citet{babarit_database_2015} and utilising the relations from Eqs.~\eqref{eq:J_reg}--\eqref{eq:capture_width}. Attenuator-type WECs usually use device length L as the normalisation length. However, to enable a fair comparison between the cases we investigate, we will use B, and later normalise by the number of buoys N.

In terms of theoretical upper limits for absorption from waves, a point absorber in combined heave and surge motion cannot exceed a capture width of $ 3\lambda/2\pi $ \citep{budal_resonant_1975,evans_theory_1976,newman_j_n_interaction_1976,jin_scalability_2022}. The CW grows with device length for attenuators \citep{jin_scalability_2022}, and can exceed 1.

\section{WEC concept and model} \label{sec:concept}

\begin{figure}[t!]
    \centering
    \includegraphics[width=0.8\linewidth]{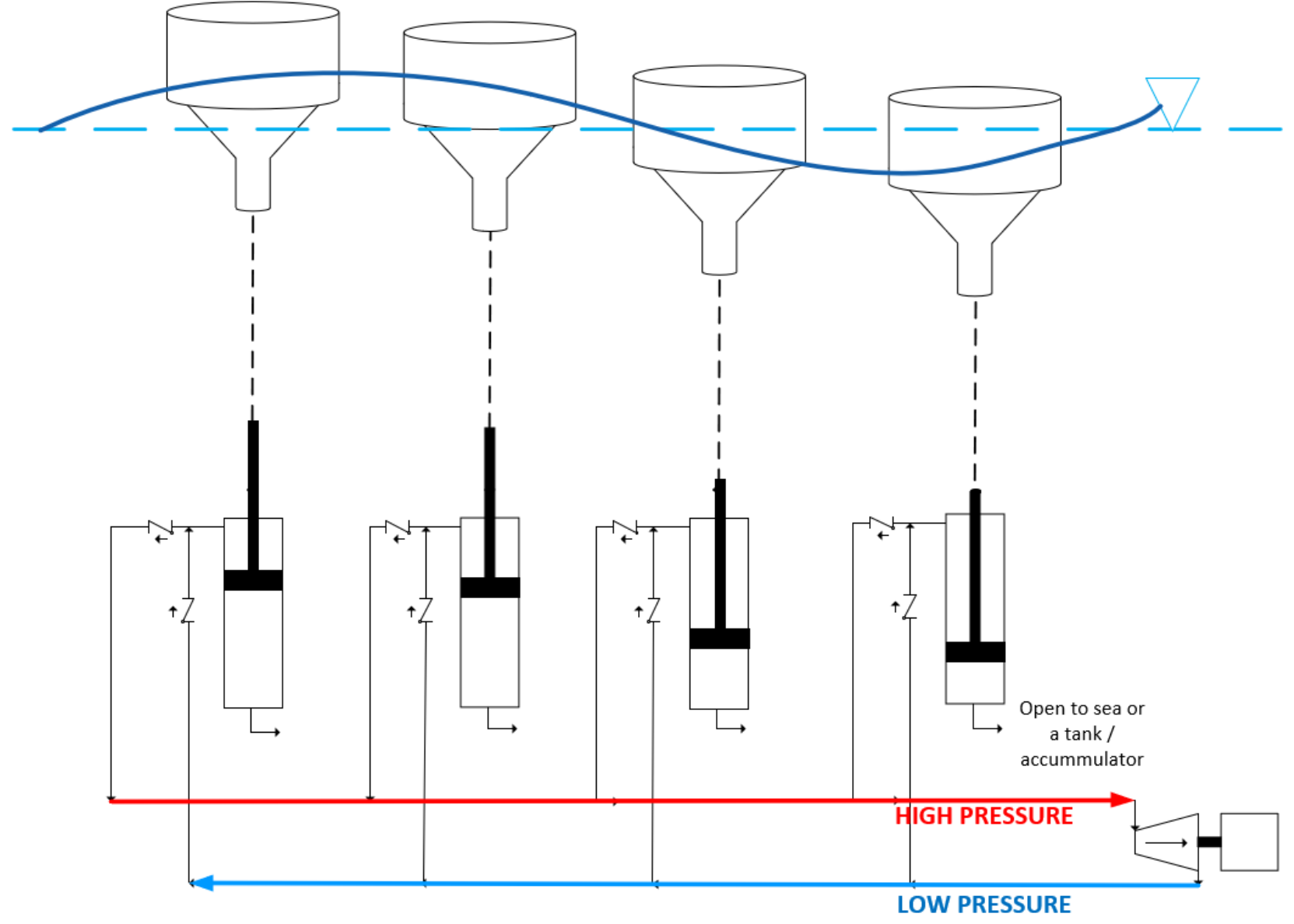}
    \caption{Schematic illustrating the working principle of the Concrest Energy WEC closed-loop hydraulic circuit. Water displaced by the upward motion of one buoy enters the high-pressure line, driving a turbine and simultaneously filling the actuator of a buoy in a wave trough, thereby assisting its descent.}
    \label{fig:concept_drawing}
\end{figure}

Concrest Energy’s Wave Energy Converter (WEC) is based on the point absorber principle, utilising surface-penetrating buoys to harvest wave energy, and is composed of multiple buoy and piston-pump modules. A schematic of the system's operating principle is shown in Fig.~\ref{fig:concept_drawing}. These modules are all integrated into a shared closed-loop hydraulic system that drives a turbine and a generator, distinguishing this system from corresponding isolated point absorbers. A buoy-type WEC is classified as a point absorber if its diameter is significantly smaller than the wavelength \citep{budal_resonant_1975}. Following recommendations from \citet{falnes_heaving_2012}, in the full-scale system, each buoy has a diameter of 10 meters and is kept in a neutral position submerged by approximately 3 meters (see Fig.~\ref{fig:buoy_sketch}). The combination of low system mass (buoys and water-hydraulics) and significant waterplane stiffness results in a system that is dynamically stiffness-dominated for prevailing wave periods, $T$, greater than 6 seconds.

Although point absorbers are recognised as the wave energy converter type with the highest power-to-size ratio, they face several functional challenges that this system aims to address. One being the return stroke of the actuator. After the actuator (typically a hydraulic pump or a directly driven generator) is activated by a wave lift, it must return to its starting position to be ready for the next stroke. This is usually achieved by gravity, a spring, or a pressurised accumulator. In a single-acting actuator, which only performs useful work during the buoy’s upward movement, the work performed when lifting the weight or tensioning the spring is lost for electricity generation. For a double-acting actuator that performs useful work in both directions, the potential energy gained by lifting the weight or tensioning the spring can be recovered as the buoy descends. Double-acting systems are, however, significantly more complex, requiring a higher component count and means to balance power extraction between strokes.

The array of point absorbers in the WEC is aligned with the direction of wave propagation. When a buoy is approached by a wave crest, the first buoy and its connected piston-rod are lifted, causing the power fluid (here: water) in the upper pump chamber to be discharged into a high-pressure line (directed by check-valves) (see Fig.~\ref{fig:concept_drawing}). The water then flows through a turbine, which extracts energy from the water and reduces the pressure, before continuing into the low-pressure line. Further, the flow will move into a pump connected to a buoy in a wave trough and retract the piston-stroke of this pump. When the turbine is allowed to rotate freely, there is negligible resistance in the hydraulic circuit, and the buoys will move in phase with the wave at high velocity but with no force (i.e., no power). As the generator is loaded, the turbine will create a resistance in the hydraulic loop and introduce a phase delay for the buoys relative to the wave. The pressures in the system will, because of the check valves upstream and downstream of the pump chamber, introduce a latching behavior. The upward stroke is latched until the pump chamber pressure exceeds that of the high-pressure line and is then discharged, while the downward stroke is latched until the pump chamber pressure drops below that of the low-pressure line, allowing the chamber to recharge. Simulations, where the turbine is modelled as a choke, show that a fixed choke cross-sectional area yields optimal energy harvesting for all wave types. Therefore, although the hydrodynamic effects were oversimplified in the simulations, the prevailing theory is that optimal capturing of wave energy coincides with the characteristics of a choke. By fortunate alignment, a variable-speed dynamic turbine exhibits a best-efficiency operating point that follows the same characteristic as a hydraulic choke, represented by the affinity line.

While resonance-type point absorbers aim to achieve high stroke through dynamic amplification at resonance, the strategy is focused on reduced stroke and higher forces for several key reasons:

\begin{itemize}
    \item The principle of operating with high pump rod force and low stroke results in high pressure and low flow within the hydraulic system. This reduced flow allows for more compact and cost-effective hydraulic equipment, while the lower hydraulic water mass enhances system responsiveness by minimizing inertia.
    \item Large stroke operation necessitates longer piston pumps, which are subject to higher stroke velocities and increased risk of impact loads at end-stops.
    \item A system that is operationally designed for high stroke must also be structurally engineered to withstand high forces during extreme weather conditions or failure modes in which motion for various reasons may be restricted (e.g., at end-stops).
    \item Resonance-operation is extremely complicated, and it can be questioned if the higher potential power capture outweighs the increased cost and vulnerability. Besides, resonance behavior may be less beneficial for large wave energy plants according to \citet{falnes_heaving_2012}.
\end{itemize}

\subsection{Experimental model}

The characteristics of the University of Oslo wave tank were taken into account when designing the test setup. The tank’s size, depth, and wave-maker system are crucial for selecting an appropriate scale. For example, the buoy diameter should be chosen to ensure it maintains the point absorber characteristic (with diameter $\ll$ wavelength) in the test environment, and the overall physical size of the system was also determined to minimize hydrodynamic interaction with the pool walls. Based on these considerations, an array of 8 buoys at a 1:40 scale was selected, corresponding to a 25 cm buoy diameter and a total length of 262.5 cm, meaning the axis-to-axis spacing between the buoys was $1.5$ diameters.

\begin{figure}[t]
    \centering
    \includegraphics[width=0.6\linewidth]{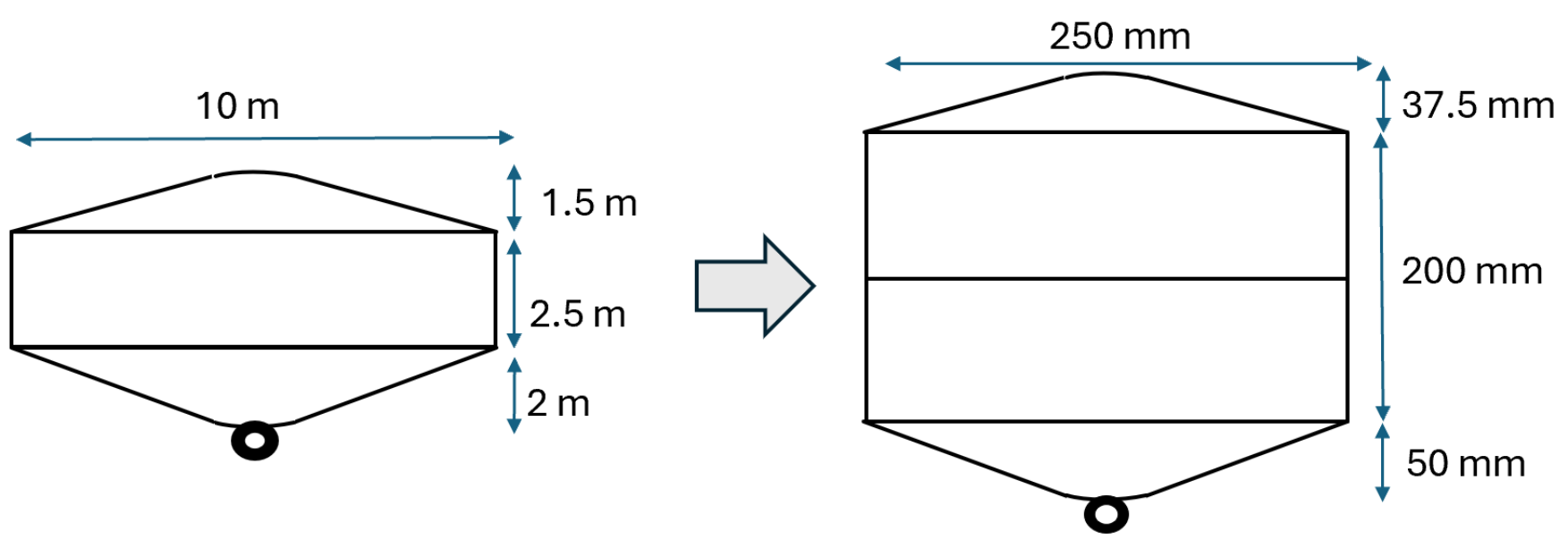}
    \caption{Dimensions of model-scale buoy and the corresponding full-scale version.}
    \label{fig:buoy_sketch}
\end{figure}

All dimensions of the buoy, except the height, were scaled 1:40. This is shown in Fig.~\ref{fig:buoy_sketch} compared to the intended full-scale version. A relatively taller buoy was chosen for the test setup for two reasons. First, since system friction becomes more dominant at smaller scales, it was a concern that testing at larger waves might be needed to obtain the desired system behaviour. Second, a taller buoy enables testing at different buoy drafts. The draft is adjusted by filling or emptying the closed loop system with water, which moves the piston rods up or down for all buoys simultaneously.

\begin{figure}[t]
    \centering
    \includegraphics[width=0.8\linewidth]{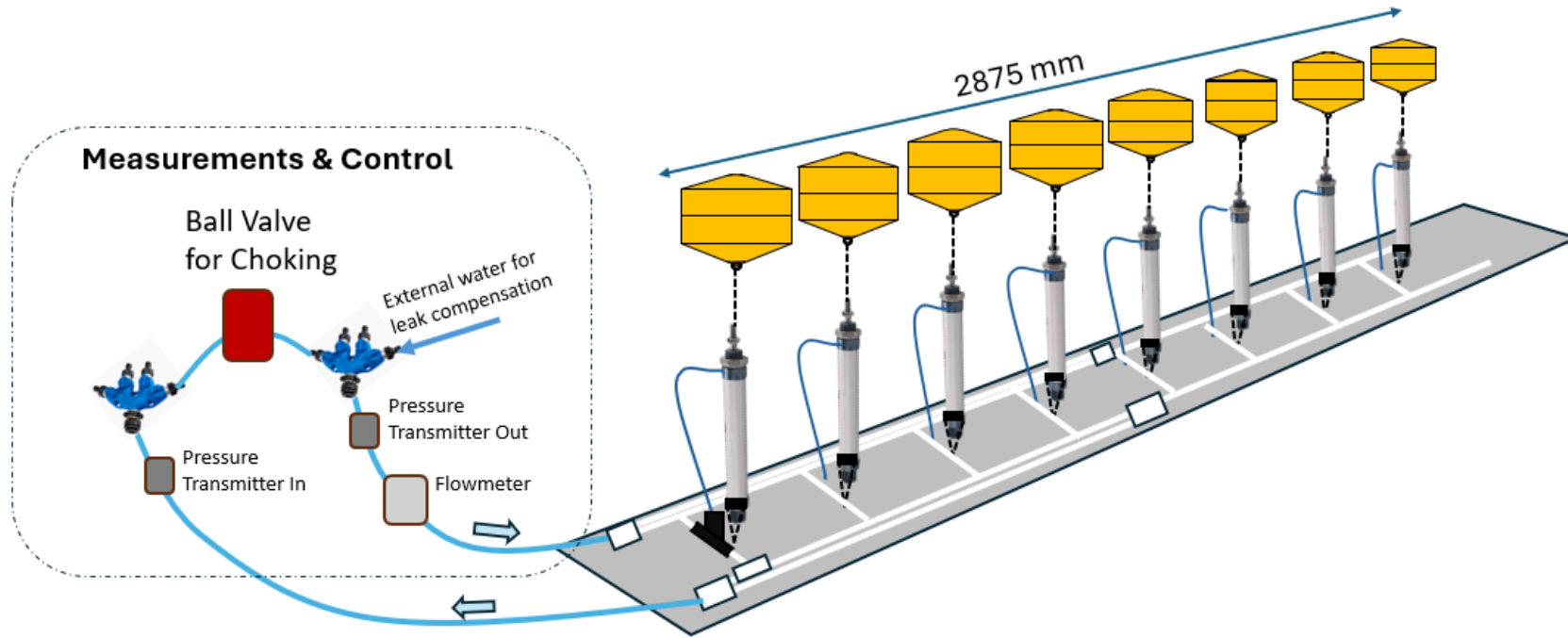}
    \caption{The model-scale Concrest Energy WEC employed in the experimental campaign, showing the placement of the pressure and flow measurement sensors as well as the choke valve. The length represents the distance between the outer edges of the first and last buoys.}
    \label{fig:model_sketch}
\end{figure}

A sketch of the experimental model WEC is shown in Fig.~\ref{fig:model_sketch}. It was deemed impractical to use the same scaling factor for the hydraulic components as for the buoys, as this would require a 10 mm pump piston diameter and 5 mm pipes, which were not readily available. Instead, piston pumps with a 24 mm piston diameter (scale 1:17) and pipes with a 10 mm inner diameter (scale 1:20) were selected. The system was fixed to the bottom of the wave tank, and hoses with a 19 mm inner diameter (scale 1:10.5) and a combined length of 540 mm connected the wave harvesting system to the control and measurement hub. The hub included a ball valve to choke the flow (representing a turbine), a flow meter (Endress+Hauser (EM), Picomag DMA20-AAACA1), and pressure sensors (Fuji Electric France S.A.S, FKCW36V5AKCYYAU) to measure pressure loss across the ball valve. To compensate for leakage in the closed-loop system, external water was supplied downstream of the ball valve.

The non-uniform scaling, combined with the long hoses to the control and measurement hub, resulted in a much larger system mass (determined by the water volume in the hydraulic system) relative to the system stiffness (determined by the waterplane area of the buoys) than would be the case with a uniformly scaled system. This reduces the system’s natural frequency and responsiveness to high-frequency waves, which will be discussed further when presenting the test data.

\section{Tank validation and experimental set up} \label{sec:validation_and_setup}

\begin{figure}[t!]
    \centering
    \includegraphics[width=0.8\linewidth]{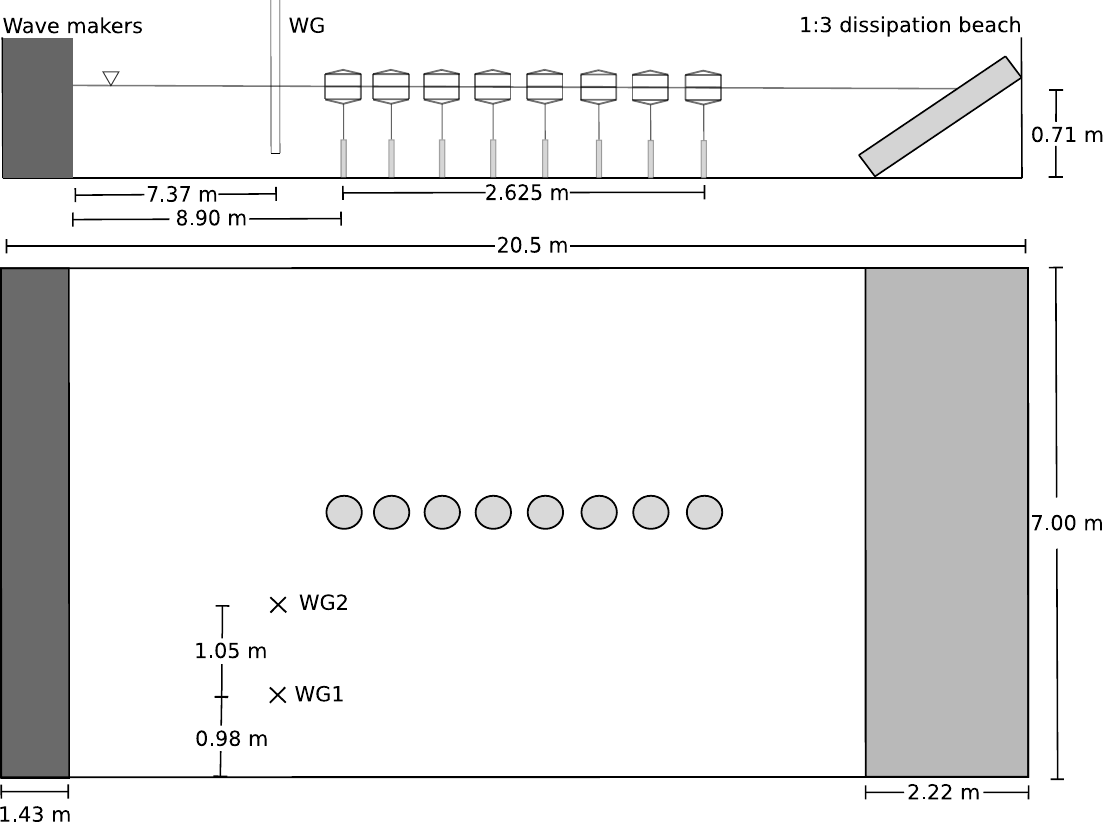}
    \caption{Schematic representation of the experimental setup used in the campaign, illustrating the WEC position in relation to the tank geometry and the location of the wave gauges. Not to scale.}
    \label{fig:experimental_setup}
\end{figure}

The experimental tests were performed in the new wave tank at the University of Oslo. The facility consists of a wave tank with an internal length of 20.5 m and a width of 7.0 m, with operational water depths ranging from 0.2 to 1.0 m. The 14 piston-type wave paddles are equipped with a force-feedback system that enables active absorption of unwanted reflected waves. The dissipation beach is a 1:3 slope metallic structure, which has been measured to provide substantial damping.

A schematic of the experimental set-up in the tank is shown in Fig. \ref{fig:experimental_setup}. It illustrates the conceptual placement of the WEC and the wave gauges relative to one another (not to scale). The wave probes are wire-resistance gauges (Edinburgh Design, WG8USB) and were placed to capture the incident wave field without being influenced by the immediate disturbance field, and slightly to the side, since all tests used long-crested waves.

\begin{figure}
    \centering
    \includegraphics[width=0.5\linewidth]{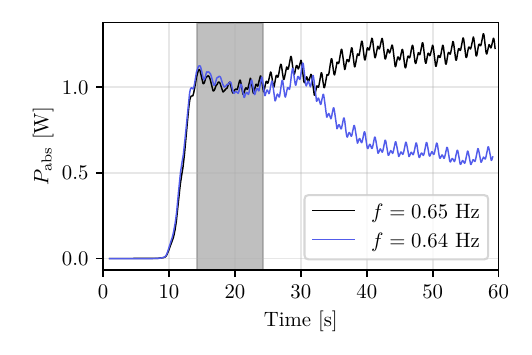}
    \caption{Comparison of the WEC response to two similar wavelength cases, highlighting the reflection effect. Incoming wave amplitude is $a=3.3$ cm for both graphs. The gray shaded region represents the analysis window for regular waves in this project.}
    \label{fig:comparingNeighbouringFrequencies}
\end{figure}

Even though the dissipation beach is a highly effective wave absorber, additional wave components had to be accounted for, which slightly constrained the data analysis for regular waves. Each buoy radiates waves in all directions, and those hitting the sidewalls will reflect as there is little to no damping in the glass walls. This causes disturbances that affect the WEC after a certain time and persist in the tank for a long time. The effect of the sidewall reflections is evident in Fig.~\ref{fig:comparingNeighbouringFrequencies}, where the incoming wave signal for the two almost similar frequencies provides similar power output in the initial phase, but diverges after the side wall reflections return, showing the sensitivity towards these reflections and their impact on the experiments.

For the data analysis of the regular-wave experiments, the following strategy was used. The start of the measurement window is determined by the group velocity of the incoming wave, $c_g$, plus an added ramp-up time from the paddles, $t_r = 8$ s, after which the signal was stable. From that point, a window of $t = 10$ s was analysed, a duration chosen somewhat pragmatically as the time before the waves radiated from the buoys return from the side walls and shown as the shaded region in Fig.~\ref{fig:comparingNeighbouringFrequencies}. Although manually selected, the overall results are not affected heavily by moderate changes in the window length, and the method is deemed robust enough.

The wave control software allows the definition of amplitude and frequency for regular waves and the construction of a target spectrum for irregular waves. We observed differences between the input amplitude and wave-gauge measurements. Therefore, all reported wave heights and amplitudes correspond to the measured values within the specified time window, measured at WG1. The input frequencies matched the measured frequencies accurately.

For optimising power output, the choke valve, described in Sec~\ref{sec:concept}, was set to the position that yielded the flow rate halfway between the fully closed and fully open position of the valve for a reference wave with amplitude 3.3 cm and $f=0.65$ Hz, from the principle that this setting yields optimal power output. Although it would have been possible to optimise the valve position for each individual case, we prioritised repeatability. The choke valve proved difficult to adjust with high precision, and maintaining a consistent setting reduced variability, as this optimisation process was also time-consuming. Accordingly, the valve was set prior to each configuration test and left unchanged unless the WEC was relocated or the buoy spacing was modified.

It is also worth noting that the flow of filling water to compensate for the piston pump leakage mentioned briefly in the Sec.~\ref{sec:concept} was injected upstream the flowmeter and hence introduced an error in the power measurement at the time of adjusting the choke. This may have caused a non-optimal choke setting. The choke setting is decisive for the buoys' latching functionality, and both measurements and simulations have shown that power is highly sensitive to the choke setting.   

The piston head leak was accounted for in the power calculation by subtracting the mean of the first 10 seconds of the time series of the flow measurements for each run. The effect of this leak and the drift of it will be discussed in Sec.~\ref{sec:regular} and \ref{sec:upscaled}. We also observed a slight offset in the pressure measurements, which was also subtracted in the same manner.

\section{Results and discussion} \label{sec:results}

During the experimental campaign, the WEC was tested in both regular and irregular wave conditions using configurations with eight and four buoys, corresponding to axis-to-axis buoy spacings of $1.5D$ and $3D$, respectively. Tests were performed with wave directions of $\theta = 0^\circ$ and $\theta = 20^\circ$, the latter achieved by rotating the WEC in the tank. We conducted a frequency sweep of regular waves from $f=0.55$ Hz to $f=0.85$ Hz ($T \approx 1.18$ s to $T \approx 1.82$ s), with each case repeated three times and a 10-minute settling period between runs. In addition, an amplitude sweep following the same strategy was carried out for $\theta = 0$, for wave steepness $ak = 0.01-0.11$.

\subsection{Regular waves} \label{sec:regular}

\begin{figure}[t]
    \centering
    \includegraphics[width=0.85\linewidth]{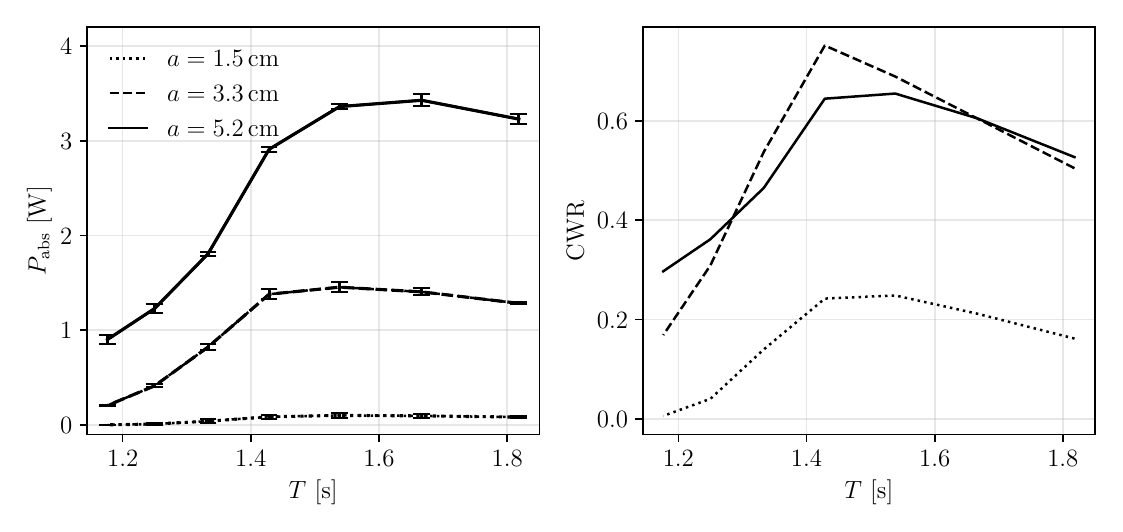}
    \caption{Power and CWR as a function of wave period, $T$ and $\theta=0^\circ$, for $a=1.5, 3.3$ and $5.2$ cm. Errorbars in a) shows one standard deviation across the three experimental runs. Only two runs were conducted for the $a=1.5$ cm case.}
    \label{fig:power_vs_frequency_fSweep_straight_8bouy}
\end{figure}

For the initial tests of the 8 buoy configuration and $\theta = 0^\circ$, we observe in Fig.~\ref{fig:power_vs_frequency_fSweep_straight_8bouy} that the system exhibits an optimal wave period range where the power output is maximised. The trend might reflect a resonant behaviour of the WEC, where such devices are generally expected to perform best \citep{mei_power_1976,scarlett_energy_2024}, and appears to be shared across multiple different WEC concepts.

\begin{figure}[t]
    \centering
    \includegraphics[width=0.5\linewidth]{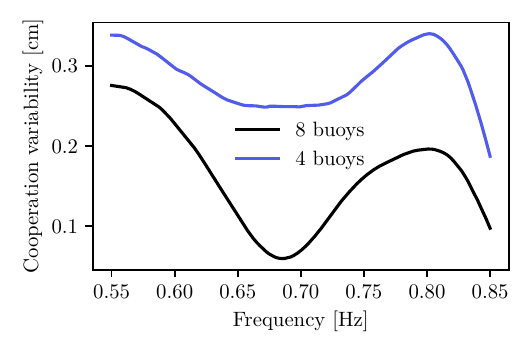}
    \caption{The standard deviation of the absolute value of the theoretical buoy displacement as a function of frequency comparing the eight-buoy and four-buoy configurations.}
    \label{fig:cooperation_variability}
\end{figure}

\begin{figure}[t]
    \centering
    \includegraphics[width=0.7\linewidth]{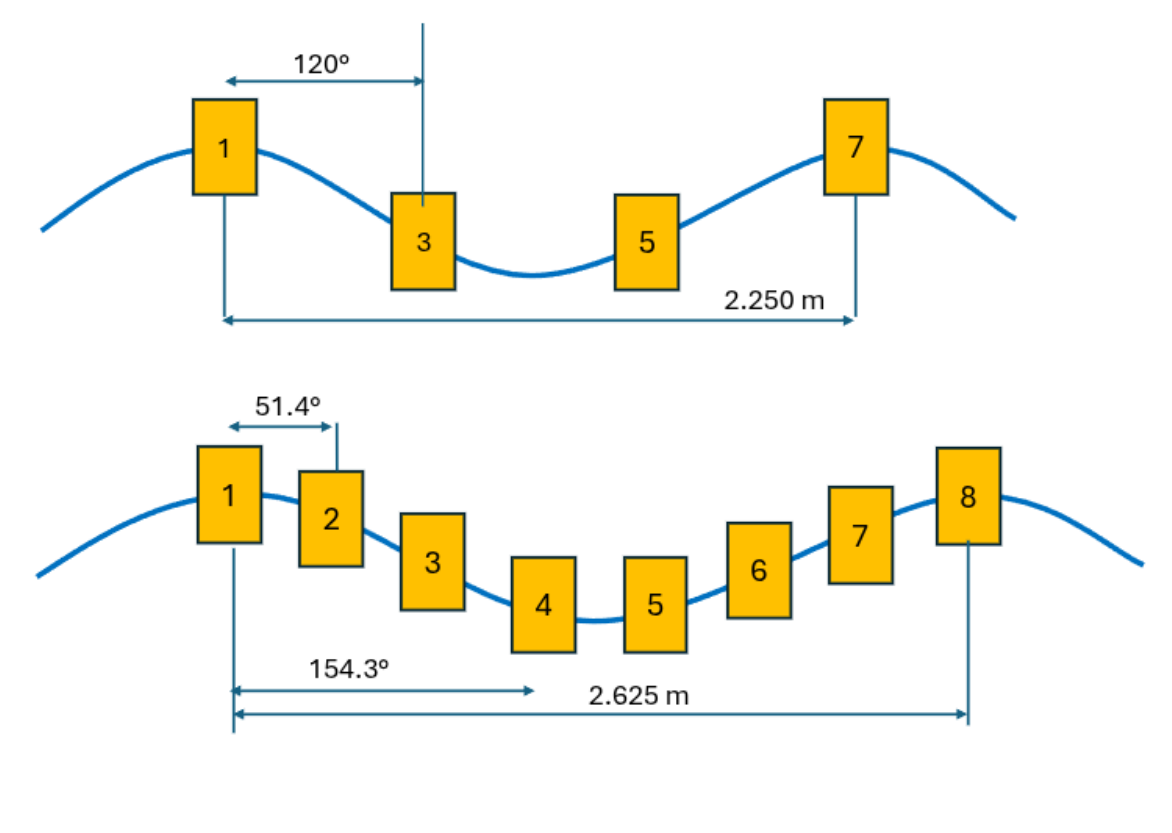}
    \caption{Comparing the two tested buoy configurations and the phase differences between cooperating buoys for wavelengths equal to the configuration length.}
    \label{fig:cooperation_ill}
\end{figure}

Another possible explanation for the optimal operating range can be illustrated by the cooperation variability shown in Fig.~\ref{fig:cooperation_variability}. In this work, we defined cooperation variability in an intentionally simplified manner: we first sum the instantaneous displacement magnitudes of all buoys, $\sum_{i=1}^{N} |z_i(t)|$, then compute the standard deviation of this time series, and divide by the number of buoys, $N$. We assume in this idealised case that all buoys perfectly follow an incoming regular wave. A low standard deviation indicates that the array's combined motion is stable and coherent. We observe that a lower cooperation variability leads to a more even flow in the hydraulic system and a smoother power output. In contrast, a high standard deviation indicates fluctuating combined motion, more uneven flow in the PTO system, and a potentially lower power output.

From the experimental time series, we observe larger fluctuations in the flow and pressure difference at $f=0.55$ Hz than at $f=0.65 - 0.70$ Hz. The trend is less clear at $f=0.85$ Hz due to buoy-buoy collisions. These collisions introduce non-periodic, chaotic behaviour in the measurements, suggesting that they are affecting the power output.

The cooperation variability of the two buoy configurations has a similar response to wave frequency, with the eight buoys slightly lower, as expected. However, around $f= 0.67$ Hz, there is a highly optimal operating frequency for the eight-buoy case that is absent in the four-buoy configuration, which is more stable as a function of frequency. We will return to the four-buoy configuration later. Although the cooperation variability is highly idealised, it provides a useful conceptual model for interpreting the results presented in the following sections.

At $f=0.67$ Hz, the wavelength corresponds to the overall system length in the 8‑buoy configuration, leading to a symmetric distribution of upward‑ and downward‑moving buoys, with each buoy finding a cooperating partner in an opposite phase. The four-buoy configuration is slightly shorter, but there is no similar dip in the cooperation variability as expected when the wavelength matches the system length at a slightly elevated frequency. The likely reason for this is that each buoy is positioned in the wave with 120 degrees spacing, far away from 180 degrees required for optimum cooperation in an idealised sine wave as illustrated in Fig.~\ref{fig:cooperation_ill}. A system with few buoys and generous inter-buoy spacing will, by design, have a high cooperation variability.  

For a system with narrowly spaced buoys, exposed to a theoretically undisturbed wave, the cooperation variability is expected to be close to zero when the system length covers an integer number of wavelengths. However, in longer arrays with more buoys, the cooperation variability will be less sensitive to a mismatch between system length and wave length, as the percentage of buoys "out of phase" will be low.

With this in mind, in Fig.~\ref{fig:power_vs_frequency_fSweep_straight_8bouy}, we observe a relatively rapid drop-off in $P_{\text{abs}}$ for wave periods shorter than the peak-power period, while the decline is more gradual for the longer waves. The response is similar for CWR, see Eq.~\eqref{eq:CWR}, and we see an opposite period response to that reported by \citet{stansby_m4_2024} for a WEC with a similar operating principle. Such differences are likely due to the unique nature of each system, model, and possibly PTO system. For our device, this response is reasonable. The reduction in absorbed power at shorter wave periods can be explained by two separate effects: system dynamics and hydrodynamic interaction. Concerning the impact of system dynamics, point absorbers become dynamically mass-dominated at wave periods shorter than the system's natural period.  Figure \ref{fig:power_vs_frequency_fSweep_straight_8bouy} indicates that our system is mass-dominated for wave periods shorter than approximately $T=1.4$ s. At these shorter wave periods, the system inertia prevents the buoys from responding adequately to the steeper, shorter-period waves. As noted in Sec.~\ref{sec:concept}, the tested model was not uniformly scaled. The combined mass of the buoys and the disproportionately large water‑hydraulic volumes implies that the effective mass of the active system is an order of magnitude higher than what a uniformly 1:40‑scaled system would exhibit. Consequently, the wave period at which the power output begins to decline would be shifted to shorter periods in the power–period diagram for a correctly scaled system. On the other hand, the hydrodynamic interactions between the buoys likely also play a role in the reduced power capture for short-period waves. For the cases with $a=3.3$ cm and $a=5.2$ cm, both buoy–buoy collisions and water slamming were observed \citep[similar to][]{xu_experimental_2019}, which likely further reduced power readings. In contrast, for the longer period, the buoys were allowed more time to respond and convert energy more effectively, and the buoys moved freely with no collisions.

%At the shorter wave periods, especially for $a=3.3$ cm and $a=5.2$, we observed both buoy-buoy collisions as well as slamming \citep[similar to][]{xu_experimental_2019}, both of which likely reduce the energy conversion efficiency. The longer waves, on the other hand, allow the buoy more time to respond and to convert energy more effectively. The behaviour may also be influenced by differences in model scaling, mentioned in Sec.~\ref{sec:concept}. The relatively large pipe diameters introduce more mass into the hydraulic loop than at the buoy scale, altering the system's resonant frequency and impairing performance at shorter wave periods.

\begin{figure}[t]
    \centering
    \includegraphics[width=0.5\linewidth]{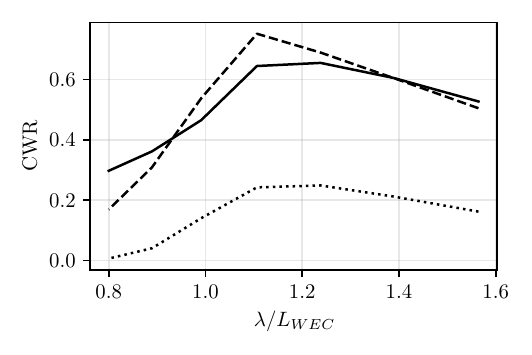}
    \caption{Capture width ratio (CWR) as a function of normalised wavelength for 8 buoys, $\theta = 0^\circ$ and $a=1.5$, $3.3$ and $5.2$ cm. $L_{\text{WEC}} = 2.625$ m being the axis-to-axis length of the WEC. Had we instead used the full length of the WEC, $L_{\text{WEC}} = 2.875$ m, the peak would have been at $\lambda/L_{\text{WEC}} = 1$. Amplitude labels as in Fig.~\ref{fig:power_vs_frequency_fSweep_straight_8bouy}}.
    \label{fig:CWR_vs_wavelength_fSweep_straight_8bouy}
\end{figure}

A similar trend is observed for the capture width ratio (CWR) in Fig.~\ref{fig:power_vs_frequency_fSweep_straight_8bouy}: there is a distinct optimal wave period, and a steeper drop-off in produced power for shorter waves compared to longer ones. An interesting observation, supporting the hypothesis that efficiency is lost due to collisions, slamming, and turbulence generation between buoys, is that the CWR for the medium-amplitude case exceeds that of the highest-amplitude waves. In an ideal, loss-free situation, we would expect the CWRs of all wave amplitudes to be similar. However, at the peak power period, the medium-amplitude waves outperform the others. At this wave period, the highest waves cause significant buoy-buoy collisions, while the medium-amplitude waves do not lead to collisions. For the shortest wave periods, there are significant collisions and slamming for both $a=3.3$ cm and $a=5.2$ cm, and more energy is extracted from the largest waves. For the smallest waves ($a=1.5$ cm), the same general period response pattern emerges, but the power extraction and CWR are much lower. The most plausible explanation for this is that friction losses and piston‑head leakage become increasingly dominant for the smaller waves, a matter to which we will return in Sec.~\ref{sec:upscaled}. In Fig.~\ref{fig:CWR_vs_wavelength_fSweep_straight_8bouy} we can see that the highest CWR aligns with a wavelength approximately the length of the WEC. This could be due to the fact that the eight buoys are more likely to find a partner in an opposite wave phase at a wavelength around the length of the WEC, especially in this configuration, as also shown by the cooperation variability.

From the reduced power output observed and with full-scale development in mind, the buoy collisions should be avoided. To explore how increased spacing could impact the system performance by reducing hydrodynamic interaction and collision avoidance, we tested a configuration in which we removed buoy 2, 4, 6, and 8 from the original model in Fig.~\ref{fig:experimental_setup}. An eight-buoy configuration with increased spacing was intended, but this would have required rebuilding the entire rig and was therefore not feasible. With only four buoys operating, the power production should be approximately reduced by 50\%. Several factors influence the performance of this slightly shorter rig compared to the eight-buoy configuration, and not all of these can be fully controlled. Nevertheless, the experiments offer useful insight into the system’s response to scaling through the addition and removal of buoys, as well as into how each configuration responds to different wave conditions. 

%We hypothesised that increasing the buoy spacing would increase robustness across wave conditions and reduce power output by less than half. 

\begin{figure}[t]
    \centering
    \includegraphics[width=0.5\linewidth]{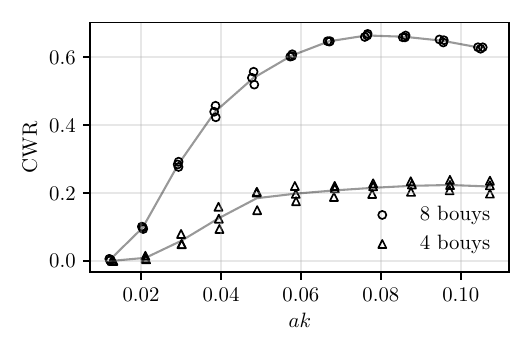}
    \caption{Capture width ratio (CWR) for the amplitude sweep with $f=0.65$ Hz comparing the two differently spaced buoy configurations.}
    \label{fig:aSweep_straight_4vs8buoy_ka}
\end{figure}

Comparing the two configurations with different spacings, we observe that the four-buoy case responds somewhat differently to the incoming waves. In the amplitude sweep presented in Fig.~\ref{fig:aSweep_straight_4vs8buoy_ka} for $f=0.65$ Hz, the CWR for the eight-buoy configuration ($1.5D$ spacing) is more than three times that of the four-buoy configuration ($3D$ spacing) at $ak \approx 0.08$. Consistent with the earlier observations, the power output is very low for the lowest steepnesses, but increases rapidly as $ak$ grows. The growth is nearly quadratic at the lower steepnesses while at higher wave steepness, however, a noticeable flattening and even a drop-off appears. For the eight-buoy configuration, this behaviour is likely related to the interaction effects, or it can be attributed to the system inertia discussed above. Interestingly, we observe the same flattening for the four-buoy configuration, despite expecting it to be less sensitive to the interaction effects, suggesting that the large system inertia is a contributor to this response. On the other hand, based on the findings by \citet{lopez-ruiz_towards_2018}, which find strong interaction effects at a spacing of $2D$ and largely minimised for a spacing of $4D$, we expect that a spacing of $3D$ still leads to significant interaction effects. Nevertheless, since we observe the CWR to be three times larger for the eight-buoy configuration, it seems that the reduced cooperation variability (i.e., improved buoy‑to‑buoy cooperation in the 8‑buoy configuration) outweighs the negative hydrodynamic interactions caused by the narrower spacing. More generally, previous studies have shown that power capture increases with the number of floats \citep{stansby_large_2017}.

%\textbf{(The conclusion of this figure must be that once the systemic friction is overcome, there is a relatively stable CWR across different wave heights for both configurations, indicating a linear increase in power, and not quadratic. Another point to be made is that interactions can be constructive and destructive, meaning that for regular waves, we could be at the frequency where this spacing is best for $1.5D$ and completely terrible for $3D$ or vice versa.)}

\begin{figure}[t!]
    \centering
    \includegraphics[width=0.85\linewidth]{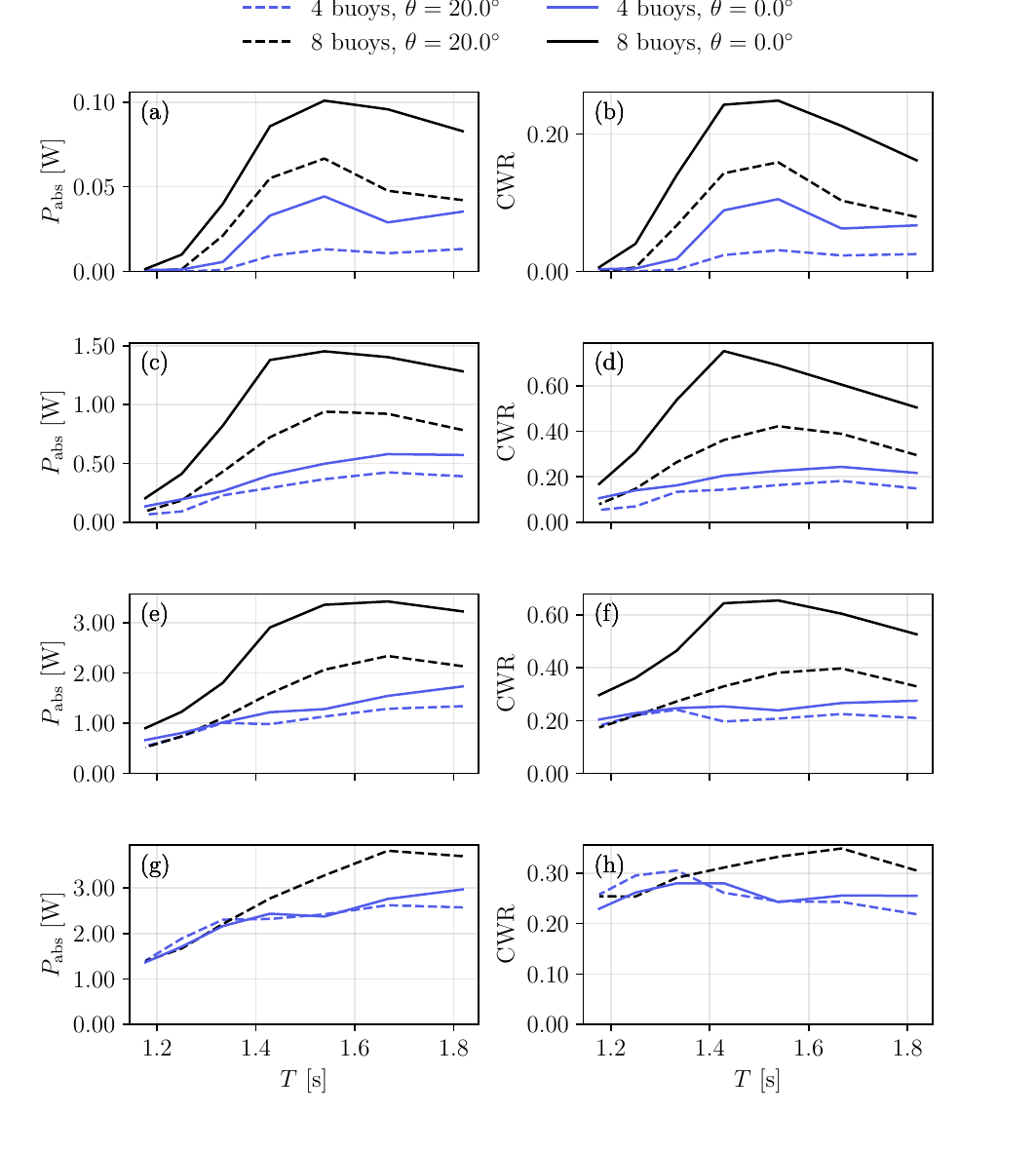}
    \caption{Absorbed power, $P_{\text abs}$, (left column) and CWR (right column) as a function of wave period for both configuration cases in $\theta = 0^\circ$ and $\theta = 20^\circ$. Amplitudes $a= 1.5$ cm (a-b), $3.3$ cm (c-d), $5.2$ cm (e-f) and $7.2$ cm (g-h). Note: difference in scales.}
    \label{fig:4x2_gigaplot_power_CWR_vs_period}
\end{figure}

Returning to the response to incoming wave period, we conducted the same frequency sweep for the four-buoy configuration, as well as for both configurations rotated $20^\circ$ in the tank. The results are presented in Fig.~\ref{fig:4x2_gigaplot_power_CWR_vs_period} for amplitudes $a = 1.5$, $3.3$, $5.2$ and $7.2$ cm. The largest-amplitude case was not tested at $\theta=0^\circ$ because the collisions observed at $a=5.2$ cm suggested that the buoys would likely be damaged if we increased the amplitude further.

Compared to the eight-buoy configuration with $1.5D$ buoy spacing, the four-buoy configuration shows a slightly different frequency response. Except for the lowest-amplitude wave case, the four-buoy configuration reaches its peak power output at longer wave periods, with $\lambda > L_\text{WEC}$. In terms of power output, the four-buoy configuration is less sensitive around the optimal wave period. This behaviour is consistent with the cooperation variability shown in Fig.~\ref{fig:cooperation_variability}, where the eight-buoy array displays a clear optimal frequency, whereas the four-buoy configuration remains comparatively stable across the tested wave periods. Particularly at shorter wave periods, the power reduction is less pronounced compared to the eight-buoy configuration, which may be explained by the avoidance of collisions and reduced destructive buoy-buoy interactions due to the larger buoy-spacing.

For both buoy-spacing configurations, we observe a reduction in power output for angled waves. This is slightly unexpected, as at $\theta=20^\circ$ the buoys should be less shadowed by one another. However, because the buoys are free to move in all modes, the angled waves cause the radiated waves from the neighbouring buoys to induce sway and roll motion. This motion might reduce the optimal wave following ability and thus energy conversion. An important note is that this effect is most prominent for the closer-spaced buoys, and we see a lower drop-off in power output for the more spaced-out array from $\theta = 0^\circ$ to $\theta = 20^\circ$. For $a=7.2$ cm, there is virtually no difference between the two wave headings for the four-buoy configuration. 

The difference in power reduction is likely influenced, in part, by drift in piston-head leakage throughout the experimental campaign. The two eight-buoy cases were tested at opposite ends of the experimental campaign, and during the test runs, we observed a gradual reduction in measured power of approximately 15-20\%. This implies that the difference in power reduction as a function of incoming wave angle is not as clear as it appears in the raw data. By contrast, the two four-buoy cases had similar leakage and can be compared directly. Although the effect of wave heading is less pronounced when the leakage is accounted for, which we will do in detail in Sec.~\ref{sec:upscaled}, a clear difference between the configurations remains. We believe this arises because, in the four-buoy configuration at $\theta = 20^\circ$, each buoy is less affected by the wake of the one ahead, whereas in the eight-buoy case, roughly half the buoy is still behind the one in front. Additionally, the four-buoy array exhibits noticeably less sway and roll motion under similar wave conditions compared with the four-buoy configuration. Further work should explore more wave headings. At some angles, it is expected that the hydrodynamic interaction will change from destructive to constructive. We hypothesize that for real irregular seastates, it will be beneficial that the waves approach the system at a certain angle, enabling energy-harvesting of a wider wave-front, as well as optimise the system to be robust towards small changes in $\theta$.

\begin{figure}[t!]
    \centering
    \includegraphics[width=0.6\linewidth]{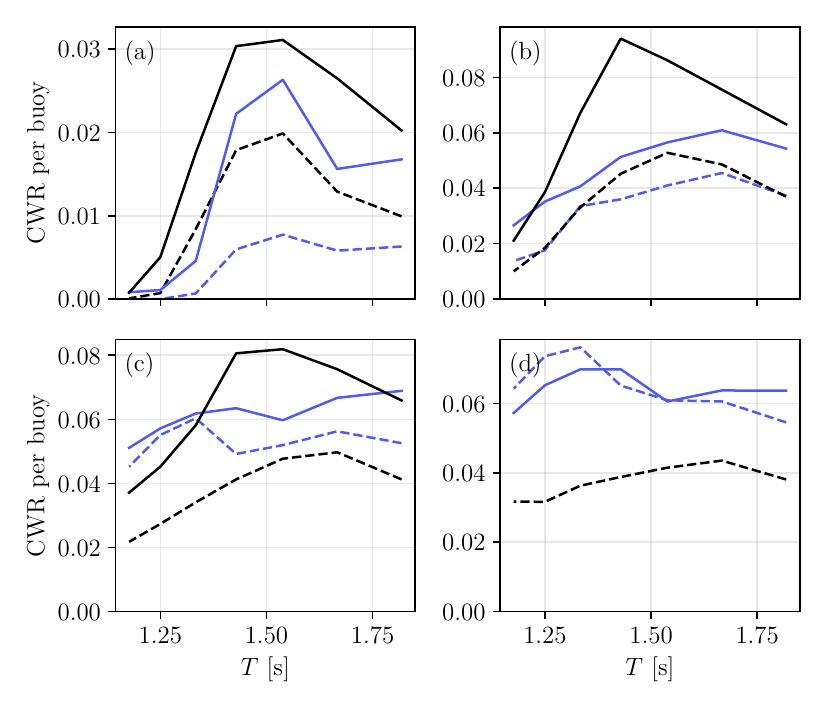}
    \caption{Buoy-normalised CWR for the four tested amplitudes, $a=1.5$ cm (a), $3.3$ cm (b), $5.2$ cm (c), and $7.2$ cm (d), comparing the two configurations in different wave headings. Note: difference in scale.}
    \label{fig:color_2x2_normalised_CWR_vs_period}
\end{figure}

The same trend is observable when comparing the CWR of the two configurations. The eight-buoy array is absorbing considerably more of the incoming wave crest than the four-buoy array, especially around $T \approx 1.4-1.5$ s. However, the four-buoy configuration appears less sensitive to frequency and exhibits a relatively stable CWR across a range of wave periods, which is a noteworthy observation. The effect of this sensitivity is clear in Fig.~\ref{fig:color_2x2_normalised_CWR_vs_period} where the CWR is normalised by the number of buoys in the array. The normalised CWR for the eight-buoy configuration is clearly more sensitive to wave period, whereas the more widely spaced array shows greater stability across incoming wave periods, especially evident at $a=5.2$ cm. This behaviour is, as mentioned, a result of the reduction of collisions and interactions between buoys and suggests the potential for the system to capture energy from a wider range of wave periods with appropriate buoy spacing. To our knowledge, such a consistently even CWR across this span of wave periods is rarely reported, especially interesting since this device adapts to incoming waves purely through its passive dynamics. As before, this observation should be interpreted cautiously, but it also represents a possible direction for the future development of this system. Even though the tests with regular waves reveal clear trends in both systemic and physical responses, we acknowledge that real irregular seas behave differently.

These observations stress the trade-off that must be addressed regarding the balance between the enhanced buoy-buoy cooperation as a result of increasing the number of buoys and the associated hydrodynamic penalties. Based on these results, the optimal system should consist of at least 8 buoys, but also include a spacing of more than $1.5D$. The most economical spacing is a balance between cost and efficiency and is a subject for further research.

%Taken together, these results may suggest that the system maintains a more robust absorption capacity over a wider range of incoming waves. Nevertheless, this interpretation should be treated with caution, as the observed reduction in destructive interaction effects relative to the eight-buoy array may also depend on other configuration-specific factors, and cannot be attributed solely to increased internal buoy spacing.

In terms of CWR per buoy, our results are somewhat lower than those of isolated point absorbers \citep[see the review by][]{babarit_database_2015}. When compared to attenuator devices, our results are slightly lower than the M4 \citep{stansby_hydrodynamics_2020,stansby_m4_2024}, similar to that observed by \citet{liao_modelling_2023} and higher than the Pelamis \citep{babarit_database_2015}. However, direct comparison is complicated by differences in device principles, model scales, and PTO systems. This is further affected by the non-uniform model scale, piston-head leakage, and manual choke settings used in the present proof-of-concept study. It is expected that the implementation of improvements identified in this study will enhance performance.

%\textbf{Note: Floating WECs, such as oscillating body systems and floating overtopping devices, operate at the water surface and capture energy through heave, pitch, or roll motions. Point absorbers and attenuators have demonstrated wide adaptability across various ocean depths and wave climates, achieving capture width ratios (CWR) up to 0.865 after optimization (Arrosyid et al., 2025). We see a CWR of 0.7-0.75 without any optimization. However, we are comparing 4/8 buoy to point absorbers etc. }

\subsection{Irregular waves} \label{sec:irregular}

\begin{figure}
    \centering
    \includegraphics[width=\linewidth]{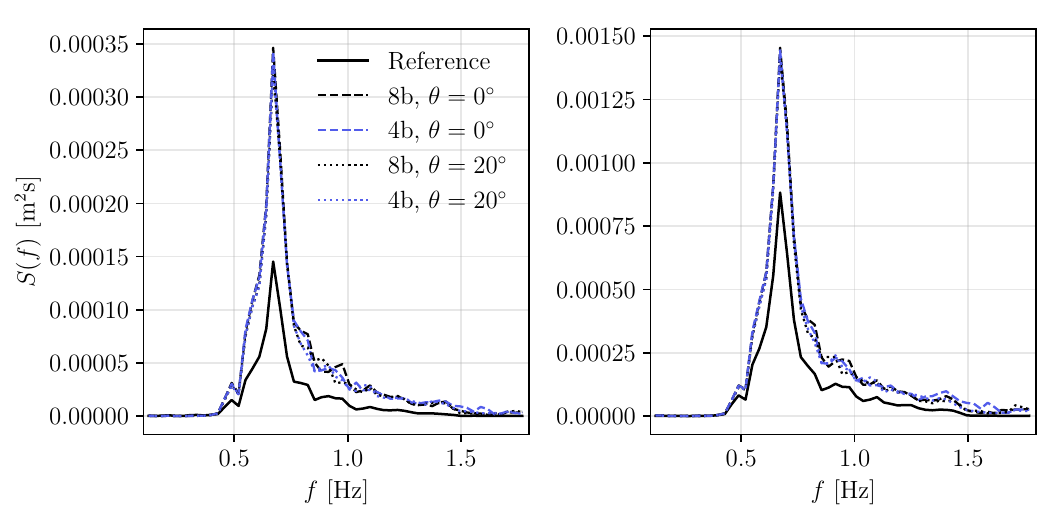}
    \caption{Spectral comparison of the four experimental irregular wave cases and the reference measurements obtained without the WEC in the tank, for $H_s = 3.2$ cm (left) and $H_s = 7.0$ cm (right).}
    \label{fig:jonswap_comparison}
\end{figure}

For the irregular wave tests, we continued to use long-crested waves. A JONSWAP-spectrum with $\gamma = 3.3$, $T_{\text{peak}}=1.43$ s as well as $f_{min} = 0.2$ Hz and $f_{max} = 1.5$ Hz was run in the tank for 5 minutes. Two significant wave heights, $H_s = 3.2$ cm and $H_s = 7$ cm, were tested. For each case, we performed three runs with a 10-minute settling period between. As before, sidewall reflections were unavoidable and substantially influenced the measured spectra, especially due to the longer duration of the experiments. Figure \ref{fig:jonswap_comparison} shows the measured spectra for the four experimental cases for each of the two significant wave heights compared with a reference spectrum obtained by running the same JONSWAP input without the WEC in the tank. Because energy propagating in the normal direction to the incoming waves is not effectively damped, it remains in the tank and increases the measured spectrum. This will impact the measured power output compared to a hypothetical case with perfectly damping walls. Whether this effect acts constructively or destructively, relating to what we observed in Fig.~\ref{fig:comparingNeighbouringFrequencies} for the regular wave cases, is not clear. What is more important for the present analysis, and clear in Fig~\ref{fig:jonswap_comparison}, is that the measured spectra are quite similar across the four experimental cases, regardless of configuration or orientation, thereby enabling a meaningful comparison of the WEC response under irregular wave conditions.

The measured significant wave heights for the reference cases were approximately $H_s = 2.0$ cm and $H_s = 5.0$ cm, whereas for the experimental runs, the measured values were around $ H_s = 3.2$ cm and $ H_s = 7.0$ cm, respectively. These were calculated using $H_s = 4\sigma$, where $\sigma$ is the standard deviation of the surface elevation time series. In the results that follow, we refer to the measured $H_s$ for the specific case rather than the reference case.

\begin{figure}[t!]
    \centering
    \includegraphics[width=\linewidth]{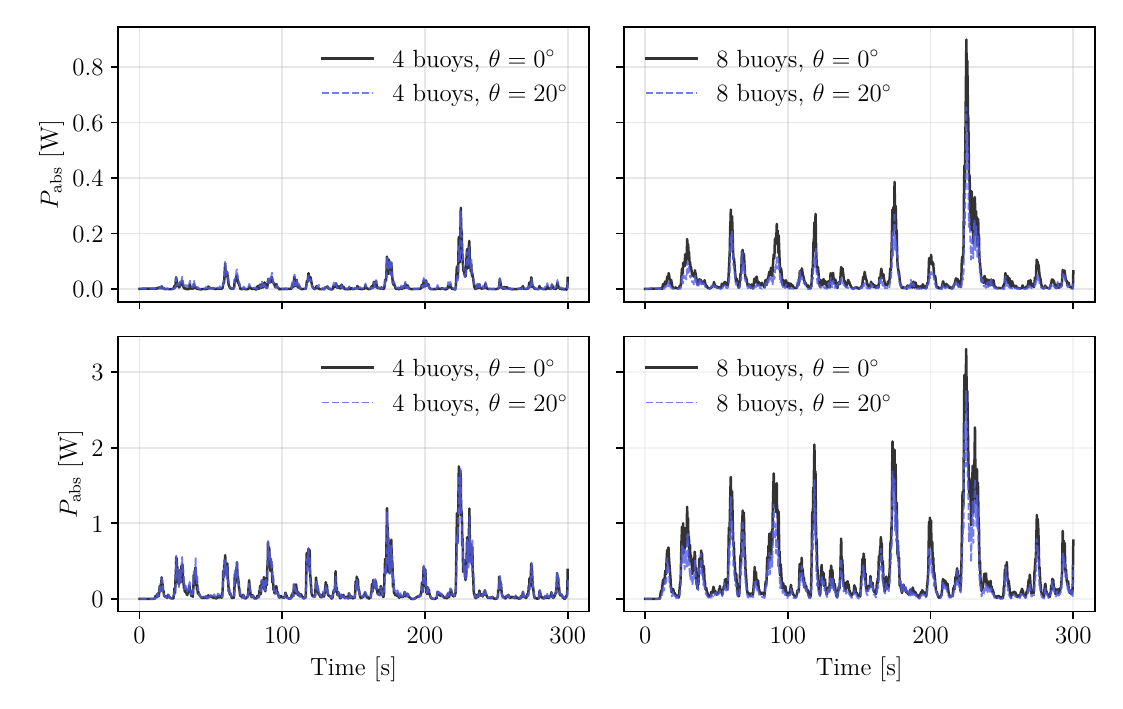}
    \caption{Time series measurements of absorbed power from the experiments with long-crested irregular waves.}
    \label{fig:2x2_JONSWAP_power_vs_time_ALLcm}
\end{figure}

The instantaneous power output measurements for the irregular wave experiments are presented in Fig.~\ref{fig:2x2_JONSWAP_power_vs_time_ALLcm}. As expected, the eight-buoy array and the higher $H_s$ case produce substantially more power. For $H_s=3.2$ cm, we observe longer periods with no power production and a very low average power output. Although smaller, the peaks occur at the same times as those of the larger waves, indicating that the WEC is responding to the same incoming wave groups. Consistent with what we have seen for $a=1.5$ cm in regular waves, the power output is very low for these small waves, likely due to system friction and mass inertia. For the four-buoy configuration and $H_s = 3.2$ cm, only 5\% of the time series contain power readings above 0.05 W, whereas this number is closer to 85\% for the eight-buoy configuration with $H_s = 7.0$ cm. These time series, with their characteristic on-off periods, are similar to what is reported in other irregular-wave experiments \citep{liao_modelling_2023,brito_experimental_2020,stansby_capture_2015,li_real-time_2018,yemm_pelamis_2012}.

\begin{table}[ht]
\centering
\caption{Summary of mean power and CWR for each configuration across the tested wave heights.}
\label{tab:mean_power_summary}
\begin{tabular}{lccc}
\hline
Configuration & $H_s$ [cm] & $P_{\text{abs}}$ [W] & CWR\\
\hline
\multicolumn{4}{l}{\textit{eight-buoy array}} \\
\hline
$\theta = 0^\circ$ & 3.27 & 0.017 & 0.089\\
$\theta = 20^\circ$ & 3.19 & 0.008 & 0.044\\
$\theta = 0^\circ$ & 7.01 & 0.245 & 0.277\\
$\theta = 20^\circ$ & 6.87 & 0.155 & 0.175\\
\hline
\multicolumn{4}{l}{\textit{four-buoy array}} \\
\hline
$\theta = 0^\circ$ & 3.25 & 0.003 & 0.014\\
$\theta = 20^\circ$ & 3.18 & 0.002 & 0.012\\
$\theta = 0^\circ$ & 7.03 & 0.078 & 0.088\\
$\theta = 20^\circ$ & 6.90 & 0.074 & 0.083\\
\hline
\end{tabular}
\end{table}

The results of these measurements are summarised in Tab.~\ref{tab:mean_power_summary}. Clearly, the average power output for the cases at $H_s = 3.2$ cm is very low, below our measurement uncertainty, but they are included for completeness and transparency. We find the highest CWR in the irregular-wave experiments for the highest waves and the eight-buoy configuration, as expected. Comparing the two buoy configurations, we still observe that the eight-buoy array absorbs more than twice as much as the four-buoy array for $\theta=0^\circ$, suggesting better cooperation between buoys. If we compare the CWR values in Tab.~\ref{tab:mean_power_summary} to the values present in the regular wave experiments, these are much lower in irregular waves across all cases. However, direct comparison between the irregular cases and the regular cases is not strictly fair. The JONSWAP-spectrum will contain a substantial amount of energy at the higher frequencies (shorter wave periods), where we have seen the WEC perform more poorly. Thus, the periods of no power production observed in Fig.~\ref{fig:2x2_JONSWAP_power_vs_time_ALLcm} are expected. Likely, the performance of the WEC in these cases will improve with lower system inertia, as previously discussed. The piston-head leakage is expected to have stolen substantial power in these cases as well, which we will address in more detail in Sec.~\ref{sec:upscaled}.

%Assuming the root-mean-square wave height to be approximately $0.7H_s$, yielding $H\approx 4.9$ cm, this value for the CWR is lower than the regular wave counterpart at $T_p = 1.43$ s, or $f=0.7$ Hz, between $a=1.5$ cm and $a=3.3$ cm. A similar trend can be observed for the three other irregular-wave cases: the absorbed power and CWR in irregular waves are clearly lower compared to the regular waves at comparable wave heights. 

However, similar to the trend observed for regular waves in Sec.~\ref{sec:regular}, there is a clear difference in how buoy spacing affects the performance of the two configurations in angled waves compared to head waves. For the eight-buoy configuration, is dropping with 37\% for $H_s = 7.0$ cm, when going from $\theta = 0^\circ$ to $\theta = 20^\circ$. Aligning with our observation in regular waves, the four-buoy configuration is more robust to changes in wave heading, and we see negligible differences in absorbed power and CWR for $\theta = 20^\circ$ compared to $\theta = 0^\circ$. This supports the suggestion that a larger buoy spacing may be more suitable for power absorption combined with a high buoy count. As a final note, these irregular-wave cases were conducted with the same system leakage rate, making direct comparison consistent.

\subsection{Addressing leakage problem} \label{sec:upscaled}

As discussed earlier, leakage from the piston heads increased throughout the experimental campaign, thereby increasing the uncertainty of the analysis. The leakage likely reduced the power absorption capability of the tested model, consistent with our measurements. \citet{falnes_budals_2003} reported a similar problem with leakage in the seals of the piston pump, clearly reducing the absorbed wave energy that was delivered to the energy store \citep{falnes_heaving_2012}.

\begin{figure}[t]
    \centering
    \includegraphics[width=0.5\linewidth]{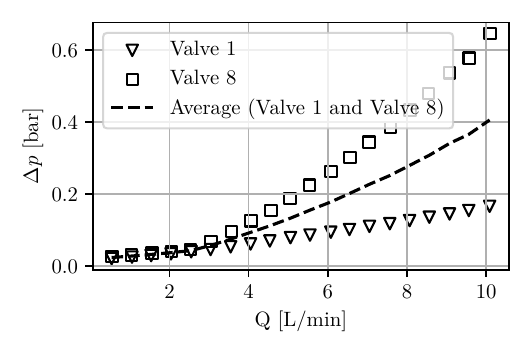}
    \caption{Pressure loss as a function of flow rate measured across valves 1 and 8, along with the calculated average pressure loss. The measurement for valve 8 includes the inlet and outlet pipes leading to and from the valve.}
    \label{fig:valve1_valve8_average}
\end{figure}

To investigate the effect of the piston head leakage, we measured pressure losses across the check valves and pipes at the flow rates observed in the experiments. The flow rate was increased in increments of 0.5 L/min and held constant for approximately 30 seconds, while pressure loss and flow rate were logged. This process was repeated twice for each of the three measured components. Figure \ref{fig:valve1_valve8_average} presents the accumulated pressure loss across the valve nearest the water supply and the valve furthest from the water supply, including the pipes. Pressure losses in the pipes were measured separately and then added due to practical constraints. The dashed line represents the average of the two measurements and serves as an estimate of the system pressure loss.

\begin{figure}[t!]
    \centering
    \includegraphics[width=0.9\linewidth]{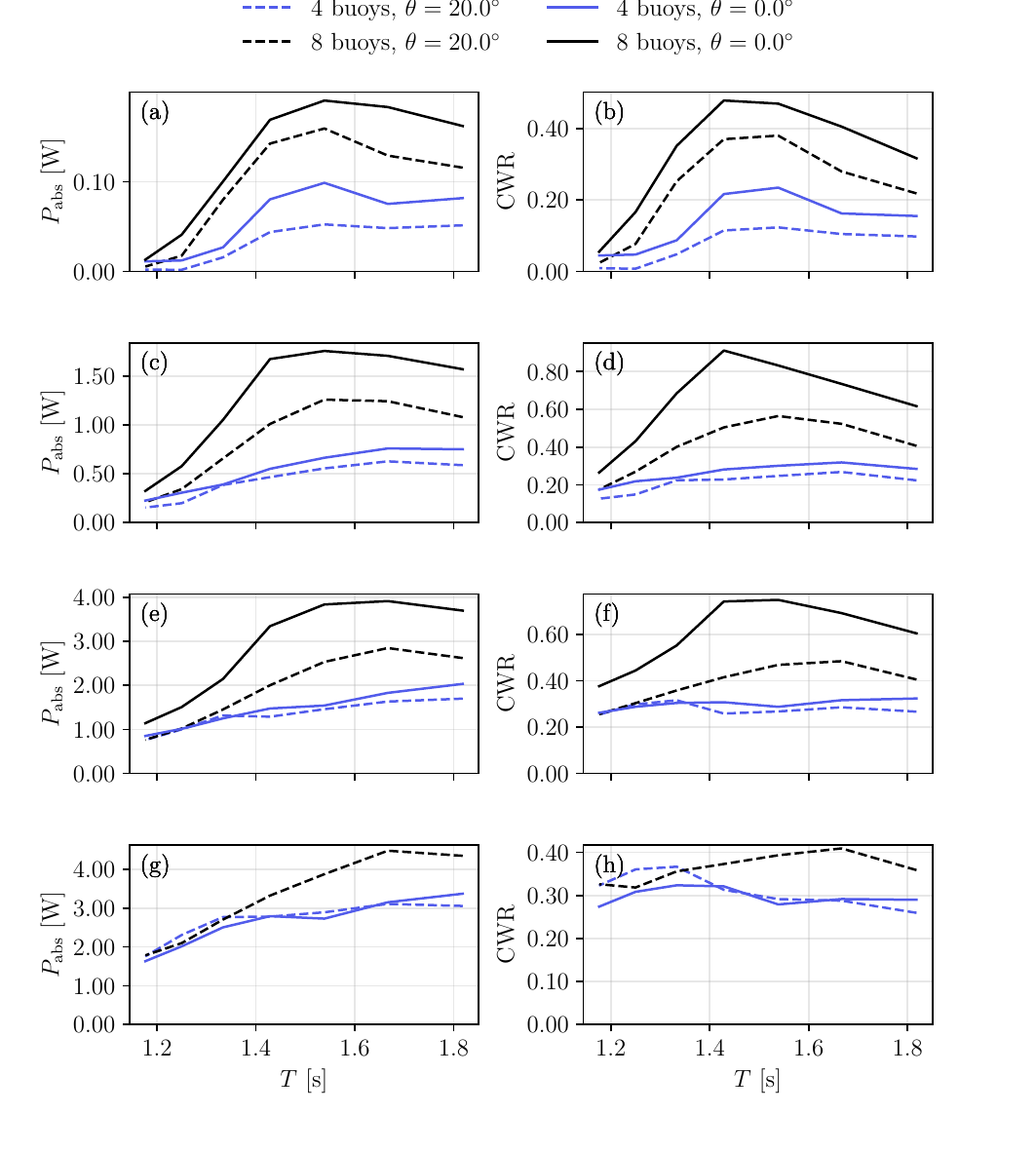}
    \caption{Absorbed power, $P_{\text abs}$, (left column) and CWR (right column) as a function of wave period for both configuration cases in $\theta = 0^\circ$ and $\theta = 20^\circ$. Amplitudes $a= 1.5$ cm (a-b), $3.3$ cm (c-d), $5.2$ cm (e-f) and $7.2$ cm (g-h). Adjusted for pressure loss caused by piston-head leakage. Note: difference in scales.}
    \label{fig:color_4x2_gigaplot_power_CWR_vs_period_frictionAdj}
\end{figure}

\begin{figure}[t]
    \centering
    \includegraphics[width=0.6\linewidth]{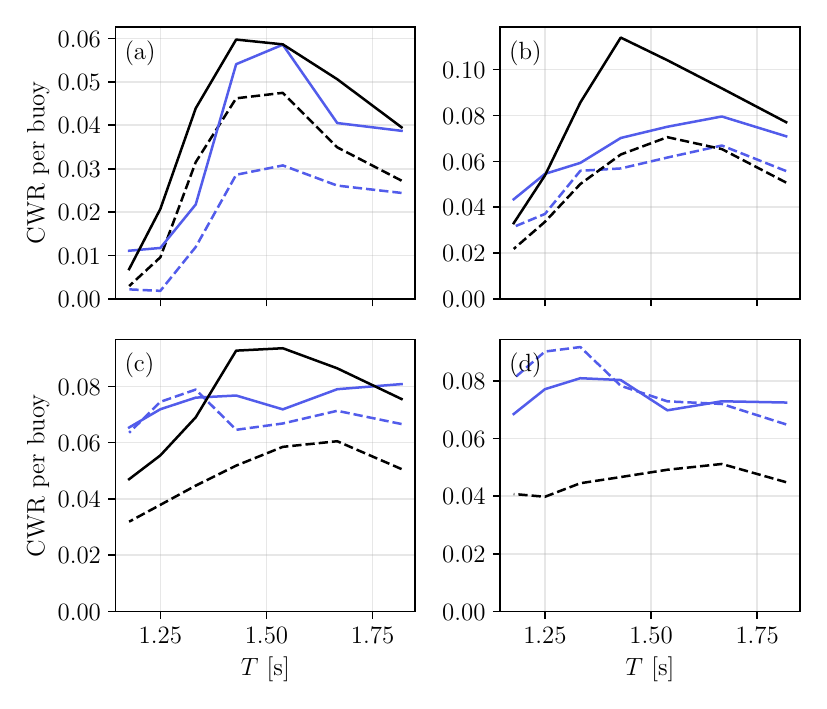}
    \caption{Buoy-normalised CWR for the four tested amplitudes, $a=1.5$ cm (a), $3.3$ cm (b), $5.2$ cm (c) and $7.2$ cm (d), comparing the two configurations in different wave headings. Adjusted for pressure loss caused by piston-head leakage. Note: difference in scale.}
    \label{fig:color_2x2_normalised_CWR_vs_period_frictionAdj}
\end{figure}

Because extra water is supplied to compensate for the leakage, the system experiences an increased internal flow and therefore additional pressure loss. To estimate the power outputs without leakage, we used the dashed line curve in Fig.~\ref{fig:valve1_valve8_average} and interpolated to obtain a function for the pressure loss as a function of the flow, $\Delta p(Q)$. The additional pressure loss introduced by the leakage is taken to be 

\begin{equation}
    \Delta p_{loss} = \Delta p(Q_{work} + Q_{leakage}) - \Delta p(Q_{work}),
\end{equation}
and the estimated power without leakage present is then
\begin{equation}
    P = Q_{work}(\Delta p + \Delta p(Q_{loss})).
\end{equation}
Based on this, we can revisit some of the main findings. Figure \ref{fig:color_4x2_gigaplot_power_CWR_vs_period_frictionAdj} shows the resulting power output and CWR after adjusting for leakage, corresponding to Fig.~\ref{fig:4x2_gigaplot_power_CWR_vs_period}. While the trends remain unchanged, the most notable effect of adjusting for leakage pressure loss is an increase in absorbed power and CWR across all cases. Upon comparison, the difference is largest for $a=1.5$ cm, which supports the conclusion that system friction, inertia, and leakage disproportionately affect the lowest amplitudes. Aside from this lowest-amplitude case, the earlier findings remain: the more widely spaced four-buoy configuration is more robust to changes in wave heading, which is also evident in the buoy-normalised CWR in Fig.~\ref{fig:color_2x2_normalised_CWR_vs_period_frictionAdj}.

For the irregular-wave experiments, Tab.~\ref{tab:mean_power_summary_frictionLoss} shows that adjusting for the pressure loss results in a clear increase in average absorbed power. The lowest wave-height cases remain very small, almost negligible. Still, the overall trend remains clear: the wider-spaced configuration is more robust to changes in wave heading for both regular and irregular waves.

\begin{table}[bht]
\centering
\caption{Summary of mean power and CWR for each configuration across the tested wave heights adjusted for pressure loss caused by leakage in piston heads.}
\label{tab:mean_power_summary_frictionLoss}
\begin{tabular}{lccc}
\hline
Configuration & $H_s$ [cm] & $P_{\text{abs}}$ [W] & CWR\\
\hline
\multicolumn{4}{l}{\textit{eight-buoy array}} \\
\hline
$\theta = 0^\circ$ & 3.27 & 0.048 & 0.246\\
$\theta = 20^\circ$ & 3.19 & 0.027 & 0.139\\
$\theta = 0^\circ$ & 7.01 & 0.389 & 0.439\\
$\theta = 20^\circ$ & 6.87 & 0.260 & 0.293\\
\hline
\multicolumn{4}{l}{\textit{four-buoy array}} \\
\hline
$\theta = 0^\circ$ & 3.25 & 0.010 & 0.049\\
$\theta = 20^\circ$ & 3.18 & 0.011 & 0.058\\
$\theta = 0^\circ$ & 7.03 & 0.130 & 0.146\\
$\theta = 20^\circ$ & 6.90 & 0.133 & 0.150\\
\hline
\end{tabular}
\end{table}

Leakage not only introduces an increased pressure loss, but also affects the overall system behaviour. When the system was newly built, initial tests revealed higher power outputs than those observed later during the experiments, even after leakage correction, at similar flow rates. This indicates that the pressure difference was the primary factor that changed and the leakage correction above does not fully account for this. This is not easily quantified, but warrants further elaboration.

The leakage is not constant when the system is in motion, but increases when the piston is lifted, and the pressure in the cylinder rises. The leakage will therefore contribute to the piston being lifted earlier and faster than it would without leakage, causing the buoy to rise more rapidly and earlier in the wave cycle. This results in a reduced buoy draught during the lifting phase and lower pressure in the piston chamber. Additionally, the leakage compensation flow (see Fig.~\ref{fig:model_sketch}) contributes to the buoy being pulled down faster into the wave troughs, increasing the pressure during the return sequence. In combination, these effects lead to a reduced pressure difference between buoy located in troughs and crests, and this mechanism supports the conclusion that $P_{\text{abs}}$ and CWR are underestimated as a result of the leakage.

Although the testing provided valuable insight into the performance in varying wave conditions and system configurations, we are cautious about scaling results to the intended full system size. Non-uniform scaling, leakage, and system losses add uncertainties concerning absolute power values and how these effects would manifest at a larger scale. Additional factors, such as mechanical losses, the choke valve positioning, different Reynolds numbers, and friction in the moving components, also contribute to efficiency losses when scaling from model to full scale.

% \textcolor{violet}{\textbf{Dette blir nok ikke med: }(As this experimental campaign had one of its main goals to inform a full-scale test, we will try to report some estimates for the full-scale power outputs of this device based on the model tests. While the physical buoy and piston size is scaled as 1:40, the hydraulic system and .. is scaled differently. Therefore these estimates must be taken with a pinch of salt. We will report the up-scaled estimates based on the 1:40-scaling (why?). The flow is free-surface and gravity-driven, such that dynamic similarity is governed primarily by the Froude number $\text{Fr} = U/\sqrt{gL}$, where $U$ and $L$ are characteristic velocity and length, respectively. By assuming Froude similarity, meaning that the Froude number for the model and the full-scale WEC is to be similar, we can estimate the full-scale power output. For the conditions tested in the irregular case, the peak period was $T_p = 1.43$ s, with $H_s = 3.2$ cm and $H_s = 7$ cm, which corresponds to $T_p = 9.04$ s, $H_S = 1.28$ m and $H_s = 2.8$ m in full scale, conditions relatively representable for sea conditions outside Smøla at the moment of writing.)}

\section{Conclusions} \label{sec:conclusion}

This study presents proof-of-concept experiments on the Concrest Energy point-absorber array WEC. The model, with a closed-loop hydraulic PTO system, was tested in both regular and irregular long-crested waves in the new wave tank at the University of Oslo. The results show that the concept absorbs incoming waves at the model scale and is capable of generating power. The closely spaced eight-buoy configuration absorbed the most power, but was accompanied by buoy collisions, slamming, and increased sensitivity to wave heading, all of which likely reduced its overall efficiency.

Increasing the internal buoy spacing and removing half of the buoys substantially altered the system behaviour. While the total absorbed power output naturally decreased with fewer buoys, the wider spacing reduced hydrodynamic interactions, minimised sway and roll excitation (visually), and produced a more uniform response across wave periods and headings. These findings suggest that destructive interaction effects within a closely spaced array may hinder the ability to absorb power across varying wave conditions, and that increasing the spacing can reduce these effects, thereby improving performance robustness.

In irregular waves, the eight-buoy array consistently absorbed more energy, primarily attributed to the improved buoy-to-buoy cooperation with a higher buoy count. The four-buoy configuration, however, demonstrated lower sensitivity to wave heading, similar to that observed in regular waves, reinforcing the advantage of wider spacing in realistic, multidirectional seas. The tests revealed reduced power capture in irregular waves, primarily due to the inability of the model WEC to absorb higher-frequency waves as a result of the non-uniform model scale and its high system mass and inertia. A uniformly scaled model is expected to improve the CWR in irregular seas.

Leakage from the piston-head seals, including drift over time, reduced the measured power absorption and introduced uncertainty into the results. Adjusting for pressure loss due to leakage increased the absorbed power output but did not alter the hydrodynamic conclusions. Nonetheless, we observe that the power output and CWR more than double from four to eight buoys, especially in irregular waves. This suggests that longer arrays with more buoys may provide greater power capture, indicating a promising direction for future scalability studies. 

%\citet{stansby_large_2017} observe a similar trend, that CWR increases with the number of floats.

Overall, this proof-of-concept study demonstrates that the Concrest Energy WEC is a promising array-based wave energy concept with potential broadband behaviour. The results indicate that performance can be enhanced by optimising internal buoy spacing and that larger arrays may yield gains in power capture and stability, but more generally that the number of buoys and internal spacing are important optimisation parameters. While uncertainties related to model scaling, leakage, choke valve setting and other unquantified mechanical losses limit direct extrapolation to full scale, the findings provide a strong foundation for further development and optimisation of the concept. Future work should focus on testing uniformly scaled models across a wider range of wave headings and buoy spacings, as well as reducing system losses, to enable more accurate assessment of full-scale performance.

% Due to the non-uniform model scaling, leakage effects, and other unquantified mechanical and hydraulic losses and effects, such as the course setting of the choke valve, which determines the latching behaviour, we are cautious about drawing definite conclusions regarding absolute power and how this would translate to a full-scale system.

% In sum, these insights provide a solid foundation for refining the concept ahead of full-scale testing and development. For further testing, a uniformly scaled model with more robust components without leakage is suggested. This model should also be designed for changing internal buoy spacing without changing the number of buoys. Further testing should also include testing of more wave headings and possibly different buoy shapes.

%% The Appendices part is started with the command \appendix;
%% appendix sections are then done as normal sections
% \appendix
% \section{Example Appendix Section}
% \label{app1}

% Appendix text.

%% For citations use: 
%%       \citet{<label>} ==> Lamport (1994)
%%       \citep{<label>} ==> (Lamport, 1994)
%%
%Example citation, See %\citet{alberello_experimental_2018}.

%% If you have bib database file and want bibtex to generate the
%% bibitems, please use
%%
%%  \bibliographystyle{elsarticle-harv} 
%%  \bibliography{<your bibdatabase>}

%% else use the following coding to input the bibitems directly in the
%% TeX file.

%% Refer following link for more details about bibliography and citations.
%% https://en.wikibooks.org/wiki/LaTeX/Bibliography_Management

%\bibliographystyle{elsarticle-harv}   % author–year style

\section*{Acknowledgements}

The authors gratefully acknowledge financial support from the UiO Growth House, University of Oslo.

\clearpage
\bibliographystyle{plainnat}
\bibliography{references.bib}

\end{document}